\documentclass[runningheads]{llncs}

\usepackage{graphicx}
\usepackage{amssymb}
\usepackage{epstopdf}
\usepackage{authblk}
\usepackage{wrapfig}
\usepackage{subfig}
\usepackage{hyperref}
\usepackage{amsmath}
\usepackage{appendix}
\usepackage{enumerate}
\usepackage{delarray}
\usepackage{stmaryrd}
\usepackage{wrapfig}
\usepackage{yfonts}
\usepackage{tikz}
\usetikzlibrary{fadings,decorations,decorations.pathreplacing}
\title{Transduction on Kadanoff Sand Pile Model Avalanches, \\ 
Application to Wave Pattern  Emergence.\thanks{Partially supported by  IXXI (Complex System Institute, Lyon) and ANR projects Subtile and MODMAD. }}
\titlerunning{Transduction on Kadanoff Sand Pile Model Avalanches}
\toctitle{Transduction on Kadanoff Sand Pile Model Avalanches}

\author{Kevin Perrot \and Eric R\'emila}
\authorrunning{K. Perrot and E. R\'emila}
\tocauthor{K. Perrot and E. R\'emila}
\institute{Universit\'e  de Lyon\\
  LIP (UMR 5668 CNRS-ENS de Lyon, Universit\'e  Lyon 1),\\
46 all\'ee d'Italie 69364 Lyon Cedex 7 - France,\\
\email{\{kevin.perrot,eric.remila\}@ens-lyon.fr }}

\begin{document}

\maketitle

\begin{abstract}
  Sand pile models are dynamical systems describing the evolution from $N$ stacked grains to a stable configuration. It uses local rules to depict grain moves and iterate it until reaching a fixed configuration from which no rule can be applied. The main interest of sand piles relies in their {\em Self Organized Criticality} (SOC), the property that a small perturbation | adding some sand grains | on a fixed configuration has uncontrolled consequences on the system, involving an arbitrary number of grain fall. Physicists L. Kadanoff {\em et al} inspire KSPM, a model presenting a sharp SOC behavior, extending the well known {\em Sand Pile Model}. In KSPM($D$), we start from a pile of $N$ stacked grains and apply the rule: $D\!-\!1$ grains can fall from column $i$ onto the $D\!-\!1$ adjacent columns to the right if the difference of height between columns $i$ and $i\!+\!1$ is greater or equal to $D$. This paper develops a formal background for the study of KSPM fixed points. This background, resumed in a finite state word transducer, is used to provide a plain formula for fixed points of KSPM(3).
    

\end{abstract}

\section{Introduction}

Sand pile models were introduced in \cite{bak88} as systems presenting a critical self-organized behavior, a property of dynamical systems having critical points as attractors. In the scope of sand piles, starting from an initial configuration of $N$ stacked grains the local evolution of particles is described by one or more iteration rules. Successive applications of such rules alter the configuration until it reaches an attractor, namely a stable state from which no rule can be applied. SOC property means those attractors are critical in the sense that a small perturbation | adding some more grains | involves an arbitrary deep reorganization of the system. Sand pile models were well studied in recent years (\cite{goles93},\cite{durandlose98},\cite{formenti07},\cite{phan08}).

\intextsep=0cm
\begin{wrapfigure}{r}{0pt}
  \begin{tikzpicture}
  \foreach \y in {1,...,3}
    \filldraw[fill=black!40] (0,.5*\y) rectangle ++ (.5,.5);
  \filldraw[fill=black!10] (0,0) rectangle ++ (.5,.5);
  \draw[dashed] (.5,0) -- ++ (.5,0);
  \draw[decorate, decoration=brace] (-.2,0) -- node [left] {$\geq D$} ++ (0,2);
  \draw[->] (.25,.75) .. controls (.75,.75) .. (.75,.25);
  \draw[->] (.25,1.25) .. controls (1.25,1.25) .. (1.25,.25);
  \draw[->] (.25,1.75) .. controls (1.75,1.75) .. (1.75,.25);
\end{tikzpicture}
  \caption{KSPM($D$) transition rule.}
  \label{fig:rule}
\end{wrapfigure}
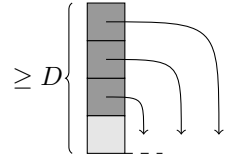

  In \cite{kadanoff89}, Kadanoff proposed a generalization of classical models closer to physical behavior of sand piles in which more than one grain can fall from a column during one iteration. Informally, Kadanoff sand pile model with parameter $D$ and $N$ grains is a discrete dynamical system, which initial configuration is composed of $N$ stacked grains, moving in discrete space and time according to a transition rule : if the height difference between column $i$ and $i+1$ is greater or equal to $D$, then $D-1$ grains can fall from column $i$ to the $D-1$ adjacent columns on the right (see figure \ref{fig:rule}).

In \cite{goles02}, the authors show that the set of reachable configurations endowed with the order induced by the successor relation has a lattice structure, in particular it has a unique {\em fixed point}.

  More formally, sand pile models we consider are defined on the space of ultimately null decreasing integer sequences. Each integer represents a column of stacked sand grains and transition rules describe how grains can move from columns. Let $h=(h_0,h_1,h_2,\dots)$ denote a {\em configuration} of the model, where each integer $h_i$ is the number of grains on column $i$. Configurations can also be given as height difference $\sigma=(\sigma_0,\sigma_1,\sigma_2,\dots)$, where for all $i \geq 0,~ \sigma_i=h_i-h_{i+1}$. We will use this latter representation throughout the paper, within the space of ultimately null non-negative integer sequences.

\begin{definition}
The   Kadanoff sand pile model with parameter $D$, KSPM($D$), is defined by:
  \begin{itemize}
    \item A set of \emph{configurations}, consisting in ultimately null non-negative integer sequences.
    \item A set of \emph{transition rules} : we have a transition from a configuration $\sigma$ to a configuration $\sigma '$ on column $i$, and we note   $\sigma \overset{i}{\rightarrow} \sigma'$ if  
    \begin{itemize}
\item $\sigma'_{i-1}=\sigma_{i-1} + D-1$ (for $i \neq 0$)
\item $\sigma'_i = \sigma_i - D$, 
\item $\sigma'_{i+D-1} = \sigma_{i+D-1} + 1$
\item $\sigma'_j = \sigma_j$ for $j \not\in  \{i-1, i, i+D-1 \}$. 
\end{itemize}
  \end{itemize}
\end{definition}

Remark that according to the definition of the transition rules, a condition for $\sigma'$ to be a configuration is that $\sigma_i \geq D$. We  note $\sigma \rightarrow \sigma'$ when there exists an integer $i$ such that $\sigma \overset{i}{\rightarrow} \sigma'$. 
The transitive closure of $\rightarrow$ is denoted by  $\overset{*}{\rightarrow}$. 
A {\em strategy} is a sequence of nonnegative integers $s=(s_1,\dots,s_T)$. We say that $\sigma'$ is {\em reached} from $\sigma$ via $s$ when $\sigma \overset{s_1}{\rightarrow} \sigma'' \overset{s_2}{\rightarrow} \dots \overset{s_T}{\rightarrow} \sigma'$, and we note $\sigma \overset{s}{\rightarrow} \sigma'$. 
We also say,  for each integer $t$ such that $0 < t \leq T$, that the column $s_t$ \emph{is fired} at \emph{time} $t$ in $s$ (informally,  the index of the sequence is interpreted as time).
For any strategy $s$ and any nonnegative integer $i$, we state $|s|_i=\#\{ t | s_t = i \}$.  Let   $s^0$, $s^1$ be  two strategies such that $\sigma \overset{s^0}{\rightarrow} \sigma^0$ and $\sigma \overset{s^1}{\rightarrow} \sigma^1$. We have the equivalence:  $[\forall~ i, |s^0|_i = |s^1|_i] \Leftrightarrow \sigma^0 = \sigma^1$.

 We say that a configuration $\sigma$ is \emph{stable}, or a \emph{fixed point} if no transition is possible from $\sigma$. A basic property of the KSPM model is the \emph{diamond property}. If there exist two distinct integers $i$ and $j$ such that 
$\sigma \overset{i}{\rightarrow} \sigma'$ and $\sigma \overset{j}{\rightarrow} \sigma''$, then there exists a configuration $\sigma'''$ such that $\sigma' \overset{j}{\rightarrow} \sigma'''$  and $\sigma'' \overset{i}{\rightarrow} \sigma'''$. 
From this property, one can easily check that, for each configuration $\sigma$, there exists a unique stable configuration, denoted by $\pi(\sigma)$, such that  $\sigma \overset{*}{\rightarrow} \pi(\sigma)$. Moreover, for any configuration $\sigma '$ such that $\sigma \overset{*}{\rightarrow} \sigma '$, we have the \emph{convergence property}: $\pi(\sigma') = \pi(\sigma)$ (see \cite{goles02} for details).

Let $N$ be a fixed integer. This paper is devoted to the structure of     $\pi((N,0^\omega))$ (where $0^\omega$ stands for an infinity of 0's. $\pi((N,0^\omega))$ is denoted $\pi(N)$, for simplification). Informally, our goal is to know what finally happens, starting from a configuration where all grains are in the first column. 

The main interest  of our approach is to provide a new tool to study fixed points: a \emph{deterministic finite state word transducer}. (a transducer is  essentially a  finite state automaton, which outputs a word during each transition).
The idea is the following, we concentrate on a set $I$ of  $D-1$ consecutive columns and constructs a set of states, according to the possible values of configurations on $I$. We then study how the way grains fall from the left part into $I$ (input word) is related to the way grains fall from $I$ to the right part (output word). The word transducer is formally defined in section \ref{subsection:transducer}. Using this transducer, an application to the case $D=3$ is also proposed in section \ref{subsection:transducer}. This application leads to: 

\begin{theorem}\label{theorem:kspm3}
  In KSPM(3), there exists a column number $i(N)$ in $O(\log N)$ such that: 
   $$\pi(N)_{[i(N),\infty[}=(2,1)^*[0](2,1)^*0^\omega$$

where $\sigma_{[i,j]}$ is the subsequence $(\sigma_i,\dots,\sigma_j)$, and $[0]$ stands for at most one 0.
\end{theorem}

This result can be interpreted as a kind of spatial emergence in a complex system. On a short length,  the structure of the  sand pile is complex, but  a very regular shape is issued from the complexity. 

Describing and proving regularity properties, for models issued from basic dynamics is a present challenge for physicists, mathematicians, and computer scientists. there exists a lot of conjectures, issued from simulations, on  discrete dynamical systems with simple local rules (sandpile model \cite{dartois} or chip firing games, but also  rotor router  \cite{levine},  the famous Langton ant\cite{gajardo}\cite{propp}...) but very few results have actually been proved.

As regards KSPM($D$), the {\em prediction problem} (namely, the problem of computing the fixed point) has been proven in \cite{moore99} to be in \textbf{NC}$^3$ for the one dimensional case (the model of our purpose), which means that the time needed to compute the fixed point is in $O(\log^3 N)$ where $N$ is the number of grains, and \textbf{P}-complete when the dimension is $\geq 3$. In this paper we give a straightforward characterization, describing asymptotically completely fixed points. A recent study (\cite{goles10}) showed that in the two dimensional case the avalanche problem (given a configuration $\sigma$ and a column $i$ on which we add one grain, does it have an influence on index $j$?) is \textbf{P}-complete, which points out an inherently sequential behavior.

\section{Avalanches and transducer}

\subsection{Previous results about avalanches}
 
Let $\sigma$ be a configuration, $\sigma^{\downarrow 0}$ is the configuration obtained by adding one grain on column $0$. In other words, if $\sigma=(\sigma_0
,\sigma_1,\dots)$, then $\sigma^{\downarrow 0}=(\sigma_0 +1 ,\sigma_1,\dots)$. Clearly,  for any strategy $s$, if $\sigma \overset{s}{\rightarrow} \sigma'$, then we have $\sigma^{\downarrow 0} \overset{s}{\rightarrow} \sigma'^{\downarrow 0}$. 
In particular, since we have: $(N, 0^\omega) \overset{*}{\rightarrow} \pi(N)$, we get: $(N, 0^\omega)^{\downarrow 0} \overset{*}{\rightarrow} \pi(N)^{\downarrow 0}$, i.e. $(N+1, 0^\omega) \overset{*}{\rightarrow} \pi(N)^{\downarrow 0}$. The model is convergent, therefore we get the inductive formula:

$$\pi(\pi(N)^{\downarrow 0})  =   \pi (N+1). $$

In the following, we will use the  inductive approach described above, which consists in computing $\pi (N+1)$ by first computing 
$\pi(N)$, then adding a grain on the origin column,  and process all possible transitions until a fixed point is reached. For initialization, $\pi(0) = 0^\omega$.

The convergence property also allows to only consider leftmost strategies. 
A strategy $s$ such that $\sigma \overset{s}{\rightarrow} \sigma'$ is called {\em leftmost} if it is the minimal strategy from $\sigma$ to $\sigma'$ according to lexicographic order. A leftmost strategy is such that at each iteration, the leftmost possible transition is performed.  
The {\em $k^{th}$ avalanche} $s^k$ is  the leftmost strategy from $\pi(k-1)^{\downarrow 0}$ to $\pi(k)$. For our iterative approach, we need to describe avalanches. 
In a previous work \cite{kspm1}, we provide a simplified description of an avalanche. This description is a core result toward the construction of the transducer in section \ref{subsection:transducer}. A first insight shows that any column is fired at most once within an avalanche. The formal statement of the avalanche description requires one more definition, even though it is graphically simple. For any sequence $x=(x_1,\dots,x_n)$ and any integers $i,j$ with $1 \leq i \leq j \leq n$, we denote $x_{[i,j]}=(x_i,\dots,x_j)$. 

Given an avalanche $s^k=(s^k_1,\dots,s^k_T)$, a column $s^k_t$ is a {\em peak} if and only if $s^k_t > \max s^k_{[1,t-1]}$. Intuitively, peaks are columns where an avalanche progresses rightward.

\begin{proposition}\cite{kspm1}\label{corollary:peak}
  Let $s=(s_1,\dots,s_T)$ be the $k^{th}$ avalanche and $(p_1\dots,p_q)$ be its sequence of peaks. Assume that  there exists a column $l$,  such that for each column  $i$ with $l \leq i < l+D-1$, $i \in s$. 
  Then for any column $p$ such that $p \geq l+D-1$, 
  $$p\text{ is a peak of }  s  \iff \pi(k-1)_p = D-1 \text{ and } \exists i \text{ s.t. } p_i < p \leq p_i+D-1$$
  
  Furthermore, let $p_i=s_t$, with $p_i \geq l+D-1$, be a peak. Then
  $$T \geq t+p_i-p_{i-1}-1 \text{ and for all } t' \text{ s.t. } t < t' \leq t+p_i-p_{i-1}-1,~ s_{t'}=s_{t'-1}-1$$

 \end{proposition}

\begin{figure}
\begin{tikzpicture}[scale=.85]
  \filldraw[fill=black] (3*.5,0) rectangle ++ (.5,.5);
  \foreach \x in {7,8,11,16,23}
    \filldraw[fill=black!30] (\x*.5,0) rectangle ++ (.5,.5);
  \foreach \x in {0,...,25}
    \draw (\x*.5,0) rectangle ++ (.5,.5);
  \draw[densely dashed] (13,0) -- ++ (.5,0);
  \draw[densely dashed] (13,.5) -- ++ (.5,0);
  \draw[densely dashed] (0,0) -- ++ (-.5,0);
  \draw[densely dashed] (0,.5) -- ++ (-.5,0);
  \draw[line width=2pt] (0,0) -- ++ (2.5,0) -- ++ (0,.5) -- ++ (-2.5,0) -- cycle;
  \node at (.25,-.25) {$l$};
  \draw[-latex] (3*.5+.25,1) -- ++ (0,-.5);
  \draw[-latex] (3*.5+.25,.5) parabola[bend pos=0.5] bend +(0,.5) +(4*.5,0);
  \draw[-latex] (4*.5+.25,.5) parabola[bend pos=0.5] bend +(0,.5) +(4*.5,0);
  \draw[-latex] (8*.5+.25,.5) parabola[bend pos=0.5] bend +(0,.5) +(3*.5,0);
  \draw[-latex] (9*.5+.25,.5) parabola[bend pos=0.5] bend +(0,.5) +(7*.5,0);
  \foreach \x in {5,6,7,10,11,13,14,15,16}
    \draw[-latex] (\x*.5+.25,0) parabola[bend pos=0.5] bend +(0,-.3) +(-.5,0);
\end{tikzpicture}
\caption{Illustration of Proposition \ref{corollary:peak} with $D=6$. Surrounded columns $l$ to $l+D-2$ are supposed to be fired. Black column is the greatest peak strictly lower than $\ell+D-1$ before the avalanche. A column is grey if and only if its value is $D-1$. Following arrows depicts the avalanche.}
\label{fig:peak}
\end{figure}
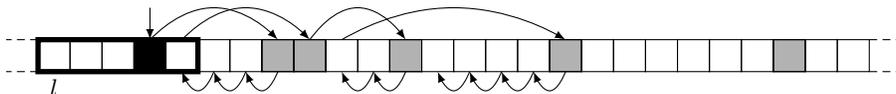

 A graphical representation of this statement is given on figure \ref{fig:peak}.

The theorem and therefore the simplified description applies starting from a certain column $l$ which depends on parameters of the model $D$ and $N$. We say that the $k^{th}$ avalanche $s^k$ is {\em dense starting at $l$ and ending at $m$} when $m$ is the greatest fired column ($\forall~ i>m,~ i \notin s^k$) and any column between $l$ and $m$ included has been fired ($\forall~ l \leq i \leq m,~ i \in s^k$). A consequence of  Proposition \ref{corollary:peak} is that the avalanche $s^k$ considered is dense starting at $l$,  where $l$ denotes the parameter in the statement of Proposition \ref{corollary:peak}. We define the {\em global density column} $L(D,N)$  as the minimal column such that for any avalanche $s^k$, with $k \leq N$, $s^k$ is dense starting at $L(D,N)$. When parameters $D$ and $N$ are fixed, we sometimes simply denote $L$.

\begin{proposition}\cite{kspm1}
The global density column $L(3,N)$ is in $O(\log N)$. 
\end{proposition}

In KSPM(3), a trivial bounding of the maximal non empty column of $\pi(N)$ shows that it is in $\Theta(\sqrt{N})$, so proposition \ref{corollary:peak} describes asymptotically completely avalanches used to construct the fixed point. We come back on this point in section \ref{subsection:conclusion}.

%%%%%%%%%%%%%%%%%%%%%%%%%%%%%%%%%%%%
%
%
%   SUCCESSIVES AVALANCHES
%
%
%
%%%%%%%%%%%%%%%%%%%%%%%%%%%%%%%%%%%%%

\subsection{Successive avalanches}

When the $k^{th}$ avalanche is dense starting at $l$ and ending at $m$, for each column $i$ such that $l + D-1 \leq i < m$, columns $i-D+1$, $i$ and $i+1$ are fired within the $k^{th}$ avalanche. Therefore, $\pi(k)_i=\pi(k-1)^{\downarrow 0}_i=\pi(k-1)_i$. Moreover, $\pi(k)_j = \pi(k-1)_j$ for $j > m+D-1$. An intuitive consequence is that two consecutive avalanches are similar. This intuition is formally stated in this section.  

Let $s^k$ denote the $k^{th}$ avalanche of KSPM($D$). We recall that the {\em global density column} $L(D,N)$ of KSPM($D$) is the minimal column such that for any avalanche $s^k$, with $k \leq N$, $s^k$ is dense starting at $L(D,N)$.
We also define $\Phi(D,N)=(\phi^1,\dots,\phi^n)$, the subsequence of $(s^1,\dots,s^N)$ reaching column $L(D,N)+D-1$. Formally, $s^k \in \Phi(D,N) \iff L(D,N)+D-1 \in s^k$. $\Phi(D,N)$ is called the {\em sequence of long avalanches up to $N$} of KSPM($D$).

 We also define the sequence $(\mu^0, \mu^1, ....., \mu^n$) of configurations such that $\mu^0 = \pi(0)  = 0^{\omega}$, and for each integer $k$ such that $\phi^k = s^m$, we have $\mu^k = \pi(m)$. 

The definition of long avalanche is motivated by the property above, which says that the effect of such an avalanche is easy to compute on the right of the global density column. 

\begin{remark}\label{remark:easy}In KSPM($D$), if $s^k$ is a long avalanche, whose sequence of peaks is denoted by $P^k$ (the largest peak being $\max P^k$), from proposition \ref{corollary:peak} we have: 

\begin{itemize}
\item $\pi(k)_{\max P^k} = \pi(k-1)_{\max P^k} -D+1 =    0$, 
\item $\pi(k)_{i } = \pi(k-1)_{i} $ for $L(D,N)+D-1 \leq i <\max P^k$,
\item $\pi(k)_{i } =  \pi(k-1)_{i} +1 $ for $\max P^k <  i  \leq  \max P^k +D -1$,
\item $\pi(k)_{i } = \pi(k-1)_{i} $ for $ i > \max P^k +D -1$.  
\end{itemize}
\end{remark}

This result is a clear application of transition rules (for each considered column $i$ we know which columns of the set $\{i-D+1, i , i+1\}$ are fired in $s^k$, so we can update the configuration). In other words, the main element that we need to compute (the right part of) $\pi(k)$ from $\pi(k-1)$ is $ \max P^k$.\\

\begin{lemma}\label{lemma:similar}
  In KSPM($D$), let $L$ be the global density column of $N$, and $\Phi=(\phi^1,\dots,\phi^n)$ its sequence of long avalanches up to $N$. Let $k < n$, and $P^k$ (resp. $P^{k+1}$) be the sequence of peaks $i$  of $\phi^k$ (resp. $\phi^{k+1}$) such that $i \geq L+2(D-1)$.  The largest peak of $P^k$ is denoted by $ \max P^k$.  
  We have: 
    $$P^k \setminus \{ \max P^k\}  ~~=~~ P^{k+1} \cap \llbracket L+2(D-1), \max P^k \llbracket$$
\end{lemma}

The lemma above can be seen as follows: $\vert P^{k+1} \vert \geq \vert P^{k} \vert -1$,  and the  $ \vert P^{k} \vert -1$ first elements of  $ P^{k+1}$ and $  P^{k}$ are equal. Informally, the  peak sequence can increase in arbitrary manner, but can decrease only peak after peak.

\begin{proof}
  Let $\kappa$, $\kappa'$ be two integers such that $\phi^k$ is the $\kappa^{th}$ avalanche, and $\phi^{k+1}$ is the $\kappa'^{th}$ avalanche. For each column $i$ such that $i \in  \llbracket L+D-1, \, \max P^k  \llbracket $, we have $i-D+1,i,i+1 \in \phi^k$ and therefore $\pi(\kappa)_i = \pi(\kappa-1)_i$.
  
  By definition of long avalanches, any avalanche $s$ between $\phi^k$ and $\phi^{k+1}$ stops before $L+D-1$, {\em i.e.} for all $i \geq L+D-1$, $i \notin s$. Combining it with previous remark, we have for all $\kappa''$ such that $\kappa \leq \kappa'' < \kappa'$
  \begin{eqnarray}
    \text{for all } i \in \llbracket L+D-1, L+2(D-1) \llbracket & , & \pi(\kappa'')_i \geq \pi(\kappa-1) \label{lemma:similar:eq1}\\
    \text{for all } i \in \llbracket L+2(D-1), \max P^k \llbracket & , & \pi(\kappa'')_i = \pi(\kappa-1) \label{lemma:similar:eq2}
  \end{eqnarray}
  because columns within interval $\llbracket L+D-1, L+2(D-1) \llbracket$ can gain height difference when a column within $\llbracket L, L+D-1 \llbracket$ is fired. This is in particular true for $\kappa'' = \kappa'-1$.
  We now study the $\kappa'^{th}$ avalanche $\phi^{k+1}$. From relation \eqref{lemma:similar:eq1} and since $\pi(\kappa'-1)$ is a fixed point, for all $i \in \llbracket L+D-1, L+2(D-1) \llbracket,~ \pi(\kappa-1)_i = D-1 \Rightarrow \pi(\kappa'-1)_i = D-1$. Let $Q^k$ (resp. $Q^{k+1}$) be the sequence of peaks $i$ of $\phi^k$ (resp. $\phi^{k+1}$) such that  $ L+D-1\leq i <  L+2(D-1)$
  Using proposition \ref{corollary:peak}, we therefore get
  \begin{eqnarray}
Q^{k}~~\subseteq~~ Q^{k+1}\label{lemma:similar:subset}
  \end{eqnarray}
  Let $I=\llbracket L+2(D-1), \max P^k \llbracket$. From relation \eqref{lemma:similar:eq2}
  \begin{eqnarray}
    \text{for all } i \in I ,~ \pi(\kappa-1)_i = D-1 \iff \pi(\kappa'-1)_i = D-1 \label{lemma:similar:eq3}
  \end{eqnarray}
  We now eventually prove the conclusion of the lemma. Let $p_{\underline i} = \min \{ i \in P^k \}$, from proposition \ref{corollary:peak} we equivalently have $p_{\underline i} = \min \{ i' \in I | \pi(\kappa-1)_i=D-1 \}$ (the existence of $p_{\underline i}$ is a hypothesis of the lemma). Let $p'_{\underline i'} = \min \{  i' \in P^{k+1}¬¨‚Ä†\} = \min \{ i' \in I | \pi(\kappa'-1)_{i'} = D-1 \}$ (the existence of $p'_{\underline i'}$ is given by subset relation (\ref{lemma:similar:subset})), using relation (\ref{lemma:similar:eq3}) we have $p'_{\underline i'} = p_{\underline i}$.\\
  Other peaks within $I$ are obviously equal from proposition \ref{corollary:peak} and relation (\ref{lemma:similar:eq2}) with $\kappa'' = \kappa'-1$.\qed
\end{proof}

%%%%%%%%%%%%%%%%%%%%%%%%%%%%%%%%%%
%
% 				TRANSDUCTEUR
%
%%%%%%%%%%%%%%%%%%%%%%%%%%%%%%%%%%%

\subsection{Transducer}\label{subsection:transducer}

\begin{wrapfigure}{r}{0pt}
\input{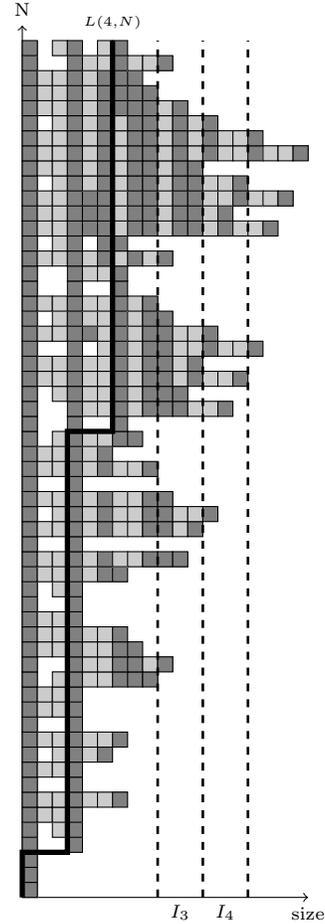}
\caption{D=4. Long avalanches up to 500, one by line. The global density column is lined in bold black. A light grey square is a fired column, a dark grey square is a peak. The sequence of types of $4$-influent subsequences up to 500 is $0,1,2,0,1,2,0,2,1,0$.}
\label{fig:av}
\vspace{-.5cm}
\end{wrapfigure}

We now exploit the similarity between successive avalanches. Informally, we will cut configurations into intervals $I_1,I_2,\dots$ of size $D-1$ and study each of them and their interactions when considering an avalanche. Given three successive intervals $I_{i-1}$, $I_i$ and $I_{i+1}$, we construct a finite state word transducer which computes the influence of $I_i$ on $I_{i+1}$, knowing the influence of $I_{i-1}$ on $I_i$ and the value of the configuration in $I_i$. The main idea to use transducers is that the value of any interval $I_i$ with $i>0$ in $\pi(0)$ is $0^{D-1}$, so we can relate temporally emergent patterns arising from transduction iterations to spatially emergent patterns on stable configurations.

 The  {\em interval} $I_i$ is the column sequence $((D-1)i, (D-1)i+1, ...,(D-1) i+ D-2))$. 
 We call {\em state} of an interval $I_i$ of a fixed point $\pi$ its value $(\pi_{(D-1)i}, \pi_{(D-1)i+1}..., \pi_{(D-1)i+D-2})$. Hence, each interval state is an element of the set  $\mathcal S = \{0,1,...., D-1\}^{D-1}$. .

We fix  an interval $I_i$  such that $(D-1)i \geq L(N)+3(D-1)$. The largest peak $j$ of $\phi^k$, such that $j < (D-1)i$, is denoted by 
 $p(i, k)$. 
 The \emph{type} $\alpha (i, k) $ of the long avalanche $\phi^k$ on $I_i$ is defined as follows. 
 \begin{itemize}
 
 \item  if $p(i, k) \in  I_{i -1}$, then $\alpha (i, k)  = p(i,k) \mod [D-1]$;
\item if $p(i, k) \notin I_{i -1}$, then $\alpha (i, k) = \epsilon$.

\end{itemize}

 Therefore, the   set of possible types is $\mathcal T= \{\epsilon, 0, 1, ..., D-2 \}$. We say that two long avalanches  are \emph{$i$-similar} if they have the same type  for $i$. Note that if  a long avalanche $\phi^k$  changes the state of $I_i$,  
 then from remark \ref{remark:easy} there necessarily exists a peak of $\phi^k$ in the interval $I_{i -1}$. 

We now divide the sequence $\Phi$ of long avalanches up to $N$ into maximal length subsequences of the type $(\phi^k,\phi^{k+1}, ...., \phi^{k''})$ such that, for each integer $k'$ such that $k \leq k' < k''$, $\phi^{k'}$ and $\phi^{k'+1}$  are $i$-similar.  Such a subsequence is called an  
\emph{$i$-subsequence}. An $i$-subsequence is said of type $\alpha$ for $i$ when the type of each  avalanche of the subsequence is $\alpha$. 
When  $\alpha$ is not the empty word $\epsilon$,  we say that the subsequence is \emph{$i$-influent}. Remark that, from Lemma \ref{lemma:similar},  each $(i+1)$-influent subsequence  is contained in an $i$-influent subsequence. See figure \ref{fig:av} for an example of $i$-influent subsequence.

\begin{lemma}\label{lemma:transducer}
Let  $\Phi_{[k,k'']} = (\phi^k,\phi^{k+1}, ...., \phi^{k''})$ be a subsequence of type  $\alpha$ for $i$, with $k'' \leq n$, and with $I_i$ an interval whose columns are greater than $L+3(D-1)$. 
Given the state $(a_0, a_1, ..., a_{D-2})$ of $I_i$ in the configuration $\mu^{k-1}$, and  $\alpha$,  one can compute, with no need of more knowledge: 
\begin{itemize}
\item the state $(a'_0, a'_1, ..., a'_{D-2})$ of $I_i$  in the configuration $\mu^{k''}$, 
\item the sequence of  types  of  the successive $i+1$-influent subsequences  contained in $\Phi_{[k,k'']}$
\end{itemize}
\end{lemma}

\begin{proof}
This is obvious when the type of the subsequence is $\epsilon$, since there is no change and the $i+1$-subsequence contained in $(\phi^k,\dots,\phi^{k''})$ is also $\epsilon$.\\
 The computation is simple when there is no integer $m$ such that $0 \leq m \leq \alpha$ and $a_m = D-1$. In this case,  the peak $p(i, k)$ is the last peak of  $\phi^k$,  thus  $\mu^{k}_{p(i, k)} = 0$, which gives that  $p(i, k)$ is not a peak of $\phi^{k+1}$, thus the subsequence is reduced to a singleton which is not $i+1$-influent (second part of the result). For $(D-1)i \leq j \leq p(i, k)+ D-1$, we have $\mu^{k}_{j} = \mu^{k-1}_{j} +1$, and for $p(i, k)+ D-1 < j <(D-1)(i+1)$, we have $\mu^{k}_{j} = \mu^{k-1}_{j}$. Thus, we have $a'_m = a_m +1 $ for $0 \leq m \leq \alpha$ and  $a'_m = a_m $ for $ \alpha < m \leq D-2$ (first part of the result).

 Otherwise, $\phi^k$ contains a peak in  $I_i$.  Let $q(i, k)$ denote the largest one. The column $q(i, k)$ is the largest $j$  such that $\mu^{k-1}_j = D-1$ and $j <(D-1)(i+1)$. Thus $q(i, k) \mod{D-1}$ is the largest $m$  such that $a_m = D-1$. 
 In this case,  $\phi^k$ starts an $i+1$-subsequence of type  $q(i, k)$. Consider the following long avalanches. From Lemma \ref{lemma:similar}, while  
$q(i, k)$ remains a peak of $\phi^{k'}$, $p(i, k)$ also remains a peak of $\phi^{k'}$. 
From Remark \ref{remark:easy}, while $q(i, k)$ is not the last peak of $\phi^{k'}$, the state of $I_i$
remains invariant. 
So the first avalanche $\phi^{k'}$ that changes the state of $I_i$ is the one whose last peak is $q(i, k)$ (this avalanche exists from the hypothesis:  $k''< n$). We have  $\mu^{k'}_{q(i, k)}  = 0$, 
which closes the $i+1$-subsequence of type $q(i, k)$. 
We also have  $\mu^{k'}_j = \mu^{k}_j +1$ for  $q(i, k) < j < (D-1)(i+1)$, and $\mu^{k'}_j = \mu^{k}_j$ for $p(i, k) \leq j < q(i, k)$. This gives the state of $I_i$ for $\mu^{k'}$  (as in the previous case, this can be rewritten to show that this state can be expressed only from $\alpha$ and  $(a_0, a_1, ..., a_{D-2})$) and proves that $p(i, k) = p(i, k'+1)$. 

The argument above can be repeated as long as we have a column $j$ of $I_i$ whose current value is $D -1$. When there is no more such column, the peak $p (i, k)$ is deleted (its value becomes 0) by the next long avalanche which is necessarily $\phi^{k''}$ from the maximality of $i$-similar subsequences.\qed
\end{proof}

The algorithm below gives the exact computation. From the state of an interval $I_i$ and an avalanche type on $I_i$, $f$ returns the greatest fired peak in $I_i$, and $g$ computes the new state of $I_i$ and appends the result of $f$ to a sequence of types on interval $I_{i+1}$. $g$ recursively calls itself, anticipating the $i$-similarity of successive avalanches when $\max P^k$ lies on the right of interval $i$.

%%%%%%%%%%%%%%%%%%%%%%%%%%%%%%%%%%%%%%%%%%%%
%
%   ALGORITHM
%
%%%%%%%%%%%%%%%%%%%%%%%%%%%%%%%%%%%%%%%%%%%%

\vspace{.2cm}
\noindent
$\left[ ~\parbox{\textwidth}{
{\small
\textbf{Input:} a non empty type $\alpha$ and an interval state $A=(a_0,\dots,a_{D-2})$.\\
\textbf{Data structure:} a sequence $w$ of types.\\
\textbf{Functions:}\\
$\begin{array}{>{\raggedright}p{.33\textwidth} | >{\raggedright}p{.6\textwidth}}
$f : \mathcal S \times \mathcal T \setminus \{ \epsilon \} \to \mathcal T$ & $g : \mathcal S \times \mathcal T \setminus \{ \epsilon \} \times T^* \to \mathcal S \times T^*$\tabularnewline
\hline
$f(A,\alpha) :=$\\
\textbf{if} $( \{ m \leq \alpha | a_m=D-1¬†\} \!\neq\! \emptyset )$\\
\textbf{then}\\
\hspace{.2cm} $\max \{¬†m | a_m = D-1\}$\\
\textbf{else}\\
\hspace{.2cm} $\epsilon$
&
$g(A,\alpha,w) :=$\\
\textbf{match} $f(A,\alpha)$ \textbf{with}\\
\hspace{.2cm} $| \epsilon \to (a_0+1,\dots,a_{\alpha}+1,a_{\alpha+1},\dots,a_{D-2}),w)$\\
\hspace{.2cm} $| p \to g((a_0,\dots,a_{p-1},0,a_{p+1}+1,\dots,a_{D-2}+1),\alpha,w\!::\!p)$
\end{array}$\\
\textbf{Computation:} $(A,\alpha) \mapsto g(A,\alpha,\epsilon)$
}
}\right.$
\vspace{.2cm}

The algorithm  above allows to define a deterministic finite state transducer \textswab T (see for example \cite{berstel}). \textswab T is a 5-tuple $(Q,\Sigma,\Gamma,I,\delta)$ where the set of states $Q$ is $\mathcal S$, the input and output alphabets (resp. $\Sigma$ and $\Gamma$) are equal to $A= T \setminus \{ \epsilon \} = \{ 0, \dots, D-2 \}$, the transition function $\delta$ has type $Q \times \Sigma \to Q \times \Gamma^*$ and is defined by the algorithm above: $\delta(q,\alpha)=Computation(q,\alpha)$. The initial state is $(0, 0,...., 0)$, and we do not need to define a final state. The image of a word $u$ by \textswab T is denoted by $t(u)$.  

\intextsep=0cm 
\begin{wrapfigure}{r}{0pt}
\begin{tikzpicture}[scale=1.5]
  \node[circle,draw,line width=1pt,fill=black!30] (21) at (0,0) {21};
  \node[circle,draw,line width=1pt,fill=black!30] (11) at (4,0) {11}
    edge [->,line width=1pt,out=-170,in=-10] node [below=-3pt] {$a|\epsilon$} (21)
    edge [<-,line width=1pt,out=170, in= 10] node [above=-3pt] {$b|ab$} (21);
  \node[circle,draw,line width=1pt,fill=black!30] (12) at (0,-4) {12}
    edge [->,line width=1pt,out=100,in=-100] node [left=-3pt] {$b|b$} (21)
    edge [<-,line width=1pt,out=80,in=-80] node [right=-3pt] {$a|a$} (21);
  \node[circle,draw,line width=1pt,fill=black!30] (22) at (4,-4) {22}
    edge [->,line width=1pt,out=-170,in=-10] node [below=-3pt] {$b|ba$} (12)
    edge [<-,line width=1pt,out=170,in=10] node [above=-3pt] {$a|\epsilon$} (12)
    edge [->,line width=1pt,out=100,in=-100] node [left=-3pt] {$a|ba$} (11)
    edge [<-,line width=1pt,out=80,in=-80] node [right=-3pt] {$b|\epsilon$} (11);
  \node[circle,draw,line width=1pt,fill=black!10] (20) at (2,-2) {20}
    edge [->,line width=1pt] node[below=-3pt] {$b|\epsilon$} (12)
    edge [->,line width=1pt] node[below=-3pt] {$a|\epsilon$} (11);
  \node[circle,draw,line width=1pt,fill=black!10] (10) at (1,-1) {10}
    edge [->,line width=1pt] node[below=-3pt] {$b|\epsilon$} (21)
    edge [->,line width=1pt] node[below=-3pt] {$a|\epsilon$} (20);
  \node[circle,draw,double,line width=1pt,fill=black!10] (00) at (2.5,-.5) {00}
    edge [->,line width=1pt] node[below=-3pt] {$b|\epsilon$} (11)
    edge [->,line width=1pt] node[below=-3pt] {$a|\epsilon$} (10);
\end{tikzpicture}
      \caption{Transducer for $D=3$ - Edges are labelled $x|u$, where $x \in \mathcal A$ is the type to the current interval (input) and $u \in \mathcal A^*$ is the resulting sequence of types applied to the next interval (output).  %For readability, $0$ and $1$ have been respectively replaced by  $a$ and $b$ in the alphabet of the transducer.  
      For example, $t(abaaaaab) = abaab$. Remark  that,  for $n >0$, we have:  $t((ab)^n) = (ab)^{n-1}$. }
   \label{fig:transducer}
\end{wrapfigure}
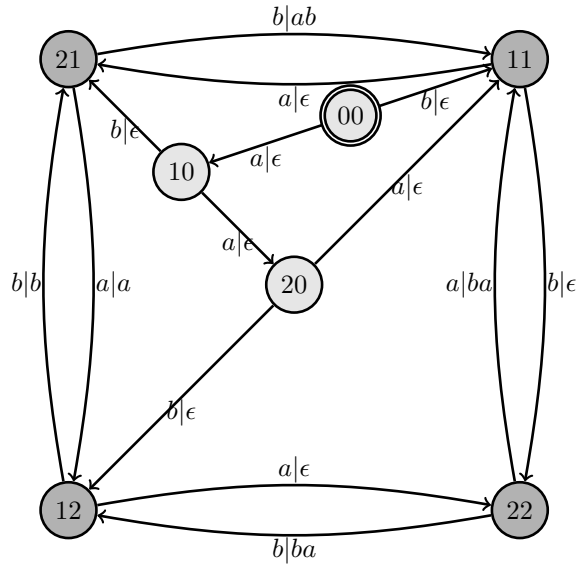

If $\alpha$ is the type of an $i$-influent subsequence for a fixed integer $i$, then  the sequence of types of the corresponding $i+1$-influent subsequences ( {\em i.e.} subsequences where considered avalanches lie) is $t(\alpha)$ . Thus, if $u$ is the sequence of types of consecutive $i$-influent subsequences for a fixed integer $i$, then $t(u)$ is the sequence of types of the corresponding $i+1$-influent subsequences. Note that the last considered avalanche may not be the last one of the last $i+1$-influent subsequence.
 
 %%%%%%%%%%%%%%%%%%%%%%%%%%%%%%
 %
 % 				D = 3
 %
 %%%%%%%%%%%%%%%%%%%%%%%%%%%%%%%%

For the lowest interesting value, $D = 3$, the transducer \textswab T can easily be drawn. 
This diagram is given  on figure \ref{fig:transducer} . For readability, we  write $a$ and $b$ instead of, respectively, $0$ and $1$,  for the alphabet of the transducer, and we omit  the drawing of states which are not connected with the initial one and are not useful for the computation of $t(u)$, for any word $u$.

The transducer has three transient states, ($00$, $10$ and $20$) and four recurrent states ($11, 12, 21$ and $22$) organized in a cycle. A non trivial analysis  of this transducer is given in appendix \ref{appendix:D=3}. The  result is stated on the lemma below. 

\begin{lemma}\label{lemma:decrease}[$D=3$]
  For any $k$ there exists $n$ in $O(\log k)$ such that for all $u$ of length $k$, $t^n(u)$ is a prefix of  $(ab)^\omega$. 
\end{lemma}

\subsection{From words to waves}

The lemma above can be used to describe fixed point configurations, noticing that $|u| \leq N$, as follows: 

\begin{proposition}\label{proposition:wave}
  In KSPM($D$), let $L$ be the global density column of $N$ and $I_i$ be an interval whose columns are greater than $L+3(D-1)$. Assume that the sequence of types of $i$-influent subsequences of long avalanches up to $N$ is
  $$(0,\dots,D-2)^x(0,\dots,p)\text{, with }x\geq 0\text{ and }p \leq D-2\text{.}$$
  Let $y$ be the size of the last subsequence of type $p$. We have $y \leq x+1$, and
  $$\pi(N)_{[i,\infty[}=\begin{array}[t]\{{ll}. (p,\dots,1)(D\!-\!1,\dots,1)^{x-y}0(D\!-\!1,\dots,1)^y0^\omega &\text{ if }y<x+1\\(p+1,\dots,1)(D\!-\!1,\dots,1)^x0^\omega&\text{ if }y=x+1\end{array}$$
\end{proposition}

\begin{proof} [sketch]
  It is a trivial induction on avalanches. We concentrate on the right part of fixed points: $\pi(k)_{[i,\infty[}$. Initially for $k=0$, it is equal to $0^\omega$. The $D-1$ first $i$-influent subsequences lead to $D-1,\dots,1,0^\omega$. Then from $(D-1,\dots,1)^x$, using lemma \ref{lemma:similar} to predict the size of each $i$-influent subsequence, we have that a sequence of type $(0,\dots,D-2)$ corresponds to exactly $(x+1)(D-1)$ long avalanches, and the behavior verifies the following invariant : the $((x+1)p+y)^{th}$ long avalanche, $0 \leq p < D-1$ and $0 < y \leq x+1$, has type $p$ and lead to
  \begin{center}\begin{tabular}[b]\{{ll}. $p,p-1,\dots,1,(D-1,\dots,1)^{x-y}0(D-1,\dots,1)^y0^\omega$ & if $y \leq x$;\\ $p+1,p,\dots,1,(D-1,\dots,1)^x0^\omega$ & if $y=x+1$.\end{tabular}\qed\end{center}
\end{proof}

\section{Conclusion}\label{subsection:conclusion}

Let us sum up results on KSPM(3). We introduced the finite state transducer which, given a sequence of types (associated with a sequence of long avalanches) on an interval $I_i$, outputs the sequence of types (associated with the same sequence of long avalanches) on interval $I_{i+1}$. We proved in a previous paper \cite{kspm1} that the global density column $L(3,N)$ is in $O(\log N)$, and therefore that toward the study of the fixed point $\pi(N)$, the word transducer applies starting from an interval $I_j$ with $j$ in $\Omega(\log N)$. Lemma \ref{lemma:decrease} shows that iterating $O(\log |u|)$ times the transducer function $t$ outputs a prefix of $(ab)^\omega$, from any input sequence $u$. An upper bound for the size of any input word (sequence of types) in KSPM(3) is $N$. As a consequence, there exists an index $k$ in $O(\log N)$ such that the sequence of types associated with subsequences of long avalanches up to $N$ on interval $I_{j+k}$ is a prefix of $(ab)^\omega$. Finally, proposition \ref{proposition:wave} converts the temporal emergence of regularities when we iterate $t$ into a spatial emergence of a wave pattern. It points out that as soon as a sequence of types which is a prefix of $(ab)^\omega$ is applied on an interval, then on the right of that interval $\pi(N)$ is a wave. A simple framing of the maximal non-empty column of $\pi(N)$ shows that it is in $\Theta(N)$, therefore the wave $(2,1)^*[0](2,1)^*$ describes asymptotically completely fixed points of KSPM(3) obtained starting from a finite number of stacked grains. This concludes the proof of Theorem \ref{theorem:kspm3}.

We hope a generalization of this result to any parameter $D$, confirming experiments:

\begin{conjecture}
For a fixed parameter $D$, there exists a column number $i(N)$ in $O(\log N)$ such that:   $\pi(N)_{[i(N),\infty[}=(D-1,D-2,\dots,2,1)^*[0](D-1,D-2,\dots,2,1)^*0^\omega$
\end{conjecture}

We name this pattern a {\em wave} for when your draw the corresponding sand pile, it looks like waves. Toward this aim, a possible outline is decomposed into two subproblems: one is to provide a general formula in $O(\log N)$ for the global density column, allowing the use of transducers from that index; a second is a general study of KSPM($D$) transducers resulting in the experimentally checked emergence of balanced outputs, then using proposition \ref{proposition:wave} we eventually conclude. Unfortunately from $D=4$, transducers lack of human-readability for their number of state is $D^{D-1}$. Nevertheless, one may  look for core properties on built transducers in order to deduce regular pattern emergence.

\bibliographystyle{plain}
\bibliography{biblio}

%%%%%%%%%%%%%%%%%%%%%%%%%%%%%%%%%%%%
%
%
%       ANNEXES
%
%
%%%%%%%%%%%%%%%%%%%%%%%%%%%%%%%%%%%%%%%%

\appendix

\section{Analysis of the transducer for $D=3$.}\label{appendix:D=3}

In this appendix we provide an analysis of the transducer for $D=3$, leading to a proof of lemma \ref{lemma:decrease}. Note that though we consider maximal length subsequences of long avalanches, input words for the transducer may contain arbitrary numbers of successive occurrences of $a$ and $b$ since we consider only $i$-influent subsequences.

We need some notations. Let $q$ and $q'$ be states of $\mathcal S$ and $u$ be a word of $\mathcal A^*$. Consider, in the transductor,  the path   which starts in $q $,  whose  sequence of successive edge (left) labels is  given by $u$. We say that we have $q\,  u = q'$ if this  path terminates in $q'$. A word $u$ is an \emph{entry} word if  $00\,  u$ is a recurrent state and for each prefix $u'$ of $u$,  $ 00\, u'$ is a transient state. We denote by $t_q$  the transduction obtained changing the initial state into $q$.  Hence $t_{00}¬†= t$. We extensively use  $t_{21}$, so we state $t_{21}¬†= t'$. A word $u$ is \emph{basic} for the state $q$ if $\vert t_q(u)\vert  \geq 2$ and for each prefix $u'$ of $u$,  $\vert t_q(u')\vert  < 2$. For each current state $q$, the set of basic words for $q$ and their images by $t_q$ are given below (trees represent case disjunctions according the beginning of $u$)

\begin{center}
  \begin{tabular}[t]{rlcl}
    $(1,1):$ & $aaaa$ & $\rightarrow$ & $aba$\\
    & $aaab$ & $\rightarrow$ & $aba$\\
    & $aab$ & $\rightarrow$ & $ab$\\
    & $ab$ & $\rightarrow$ & $ab$\\
    & $ba$ & $\rightarrow$ & $ba$\\
    & $bb$ & $\rightarrow$ & $ba$
  \end{tabular}
  \begin{tabular}[t]{rlcl}
    $(2,1):$ & $aaa$ & $\rightarrow$ & $aba$\\
    & $aab$ & $\rightarrow$ & $aba$\\
    & $ab$ & $\rightarrow$ & $ab$\\
    & $b$ & $\rightarrow$ & $ab$
  \end{tabular}
  \begin{tabular}[t]{rlcl}
    $(1,2):$ & $aa$ & $\rightarrow$ & $ba$\\
    & $ab$ & $\rightarrow$ & $ba$\\
    & $ba$ & $\rightarrow$ & $ba$\\
    & $bbu$ & $\rightarrow$ & $ab$
  \end{tabular}
  \begin{tabular}[t]{rlcl}
    $(2,2):$ & $a$ & $\rightarrow$ & $ba$\\
    & $b$ & $\rightarrow$ & $ba$
  \end{tabular}
\end{center}

Each word $u$ (such that $t(u)  \neq \epsilon$) admits a unique decomposition $u = u_0 u_1 ...u_p$ such that $u_0$ is an entry word, for $1 \leq i < p$,  $u_i$ is a basic word for the state $00\,  u_0 u_1 ...u_{i-1}$, and $u_p$ is a non-empty prefix of a basic word (for the state $00 \, u_0 u_1 ...u_{p-1}$). The word $u$ also admits a decomposition $u =u'_1 u'_2 ...u'_{p'}$ such that for $1 \leq i < p$,  $u_i$ is a basic word for the state $21\,  u_0 u_1 ...u_{i-1}$, and  $u'_{p'}$ is a non-empty prefix  of a basic word (for the state $21 \, u'_0 u'_1 ...u'_{p'-1}$).

A first result gives us a hint on the form of the sequence of types  applied to successive intervals:

\begin{lemma}\label{lemma:abu}

Let $\mathcal L$ be the language $\mathcal L = \{ab u,  u \in \mathcal A^* \} \cup   \{\epsilon, a\}$.   
\begin{itemize}
\item For each $u \in \mathcal A^*$,  we have  $t'(u) \in \mathcal L$. 
\item For each $v \in \mathcal L$,  we have  $t(v) \in\mathcal  L$.
\item For each $u \in \mathcal A^*$ , we have  $t^{2}(u) \in \mathcal L$
\end{itemize}
\end{lemma}

\begin{proof}
We prove the three items successively, using previous ones as hypothesis. 
\begin{itemize}
\item
Let $u \in \mathcal A^*$ such that $u \neq \epsilon$. Consider the second decomposition seen above: 
$u =u'_1 u'_2 ...u'_{p'}$. We obtain $t'(u) =  t'(u'_1) t_q(u'_2 ... u'_{p'})$, where $q$ denotes a recurrent state. 

\begin{itemize}
\item
For $p' \geq 2$,   $t'(u'_1)$ is the image of a basic word for $21$,  thus
$t'(u'_1) \in \{ab, aba\}$, which gives $t'(u) \in \mathcal L$. 
\item For $p' = 1$, $t'(u) =  t'(u'_1)$ and $t'(u'_1)$ is the image of non empty prefix a basic word for $21$, thus $t'(u'_1)$ is a prefix of $aba$, which gives $t'(u) \in \mathcal L$. 
\end{itemize}

\item Let $v  \in \mathcal L$. If  $v \in \{\epsilon, a\}$, then $t(v) = \epsilon$. Otherwise $v$ can be written $ab u$. Thus $t(v) = t(ab) t'(u) = t'(u)$, and  $t'(u) \in \mathcal L$ from the first item. This proves: $t(v) \in\mathcal  L$. 

\item Let $u \in \mathcal A^*$ such that $u \neq \epsilon$.  We consider the first decomposition above:  $u = u_0 u_1 ...u_p$.  We obtain $t(u) =  t_q(u_1) t_{q u_1} ( u_2...u_p)$, where $q$ denotes a recurrent state. 

\begin{itemize}
\item
For $p = 0$,   $t(u) = \epsilon$,  thus $t^{2}(u) = \epsilon$.  
\item For $p = 1$,   $t(u) = t_q(u_1)$, and $t_q (u_1)$ is  the image by $t_q$ of a
prefix of basic word for $q$,  which gives that $t(u)$ is a prefix of either $aba$ or $ba$  (since possible images of basic words are $ab, ba$, and $aba$). This gives that $t^2(u) \in \{\epsilon, a\}$. 
\item If $p \geq 2$,  then $t_q(u_1) \in \{ab, ba, aba\}$. If  $t_q(u_1) \in \{ab, aba\}$, then $t(u) \in \mathcal L$, thus $t^2(u) \in \mathcal L$, from the second item. If  $t_q(u_1) = ba$, then we can state $t(u) = ba u'$. Thus $t^2(u) = t'(u')$.  We have  
$t'(u) \in \mathcal L$ from the first item, thus $t^2(u) \in \mathcal L$. 
\end{itemize}
\end{itemize}\qed
\end{proof}

\begin{definition}[Height]
  The height $h$ of a finite word $u \in A^*$ is $h(u)= \vert |u|_a - |u|_b \vert $ where $|u|_x$ is the number of occurrences of the letter $x$ in $u$.
\end{definition}

\begin{lemma}
  For any finite word $v \in \mathcal L$, we have: $h(t(v)) \leq \frac{h(v)}{4}+1$
\end{lemma}

\begin{proof}
This is obvious if $v \in \{\epsilon, a\}$. Thus, stating $v = abu$, it remains to prove that, for any finite word $u \in \mathcal A^*$, we have: $h(t('u)) \leq \frac{h(u)}{4}+1$. 
 
  Let us first consider the case when $|u|_a - |u|_b \geq 0$. Assume that  we  remove a  pattern  of the form $ab$ or $ba$ from $u$. This does not change the value of $h(u)$. Moreover, 
   for each recurrent state $q$, $t_q(ab)$ and  $t_q(ba)$ both are elements of $\{ab, ba\}$ and $q ab = q ba = q$. This  guarantees that  pattern suppression does not change the value of $h(t'(u))$.  
  
  Iterating this argument until there is no more pattern as above leads to the following fact:    if we state $u'=a^{h(u)}$, then we have  $h(t'(u'))=h(t'(u))$.
  
   The integer $h(u)$ can be written as $h(u)=4i+r$, with $0\leq r \leq 3$. We have: $t'(aaaa) = aba$,  and $21 aaaa = 21$. Thus   
   $t'(u') = (aba)^i \,t'(a^r)$, which gives   $h(t'(u')) \leq h((aba)^i)+h(t(r)) \leq i+1$.  Thus    $h(t'(u)) \leq \frac{h(u)}{4}+1$.

The other case, when $|u|_a - |u|_b \leq 0$,  is similar. By simplifications of factors $ba$ and $ab$,  we obtain that $h(t'(u'))=h(t'(u))$, for $u'=b^{h(u)}$. The value $h(u)$ can be written as $h(u)=4j+s$, with $0 \leq s \leq 3$. We have:  $t'(bbbb) = abbab$ 
and $21 bbbb = 21$. Thus   
   $t'(u') = (abbab)^j \,t'(b^s)$, which gives   $h(t'(u')) \leq h((abbab)^j)+h(t(s)) = j$. Thus  $h(t'(u)) \leq \frac{h(u)}{4}+1$.\qed
\end{proof}

\begin{corollary}\label{corollary:decrease}
  Given a word $u \in \mathcal A^*$ of length $l$, there exists an $n(l)$ in $O(\log{l})$ such that $t^{n(l)}(u)$ is a prefix of $(ab)^\omega$.
\end{corollary}

\begin{proof}

We first prove it restricting ourselves on words of $\mathcal L$
  Given a finite word $v$ on $\mathcal L$, we define the maximal height $g(v)= \max \{ | h(v') | ~| v' \text{ prefix of }v\}$. 
  The previous lemma gives the result $g(t(v)) \leq 1 + \frac{g(v)}{4}$. We can now use a trick to get the expected result. We define $g'(v)=g(v)-\frac{4}{3}$, then: 

 $$g(t(v)) \leq 1+\frac{g(v)}{4} \iff g'(t(v)) \leq \frac{g'(v)}{4}$$

 From  lemma  \ref{lemma:abu}, $t(v)$ is element of $\mathcal L$. Thus we can iterate the inequality.  By this way, we obtain,  for each positive integer $n$: 
$$ g'(t^n(v)) \leq \frac{g'(v)}{4^n}$$

Thus, for $n>\log_4 (g'(v)) - \log_4 (\frac{2}{3})$, we have:  $g'(t^n(v)) < \frac{2}{3}$,  so $g(t'^n(v)) < 2$ and, by integrity,  

$$g(t^n(v)) \leq 1$$

This last inequality enforces that  $u$ admits a decomposition $t^n(v) = w_1, w_2  ...  w_q$ such that, for $1 \leq i < q$,   
$u_i \in \{ab, ba\}$, and  $w_q \in  \{\epsilon, a, b\}$. 
Thus $t^{n+1}(u) = t(w_1) t'(w_2),  ......t'(w_q)$.  Thus, $t^{n+1}(u)$ is a prefix of the infinite word $(ab)^\omega$, since $t'(ab) = t'(ba) = ab$ and $t(ab) = t(ba) = \epsilon$. 

Now,  if we take a finite word $u$ on $\mathcal A^*$, we have, from  lemma  \ref{lemma:abu}, $t^2(u) \in \mathcal L$. On the other hand,   
 $\vert t^2(u)   \vert \leq 4 \vert u   \vert$and $\vert t^2(u)   \vert +\frac{4}{3} \geq g'(t^2(u))$, which gives  $g'(t^2(u)) \leq 4 \vert u   \vert +\frac{4}{3}$. Therefore, for 
for $n>\log_4 (4 \vert u   \vert +\frac{4}{3}) - \log_4 (\frac{2}{3})$,  we obtain that  $t^{n+1}(t^2(u))$ is a prefix of the infinite word $(ab)^\omega$. In other words,  for $m  >\log_4{(4\vert u   \vert+\frac{4}{3})}¬†- \log_4 (\frac{2}{3}) +3$, $t^{m}(u)$ is a prefix of the infinite word $(ab)^\omega$.\qed

\end{proof}

Let us remark that $l<N$ for any input word $u$ so corollary \ref{corollary:decrease} apply for actual sand pile behavior. We therefore have a strong property on words emerging from iterations of the transducer function $t$ : they are exponentially quickly prefixes of $(ab)^\omega$.

\end{document}